\documentclass[9pt,twocolumn,twoside]{pnas-new}
\templatetype{pnasresearcharticle}
\usepackage{epsfig,latexsym}
\usepackage{amsmath,amscd,amssymb,color}
\usepackage{ulem}
\usepackage{makecell}

\newcommand{\be}{\begin{equation}}
\newcommand{\ee}{\end{equation}}
\newcommand{\bea}{\begin{eqnarray}}
\newcommand{\eea}{\end{eqnarray}}

\makeatletter
\renewcommand*{\p@subsubsection}{\thesection.}
\makeatother

\def\c#1{\textcolor{blue}{#1}}

\begin{document}

\title{Physical modeling of ribosomes along messenger RNA: 
estimating kinetic parameters from ribosome profiling experiments using a ballistic model}

\author[a]{Carole Chevalier} 
\author[a]{J\'er\^ome Dorignac}
\author[a,b]{Yahaya Ibrahim}
\author[c]{Armelle Choquet}
\author[c]{ Alexandre David}
\author[d]{Julie Ripoll}
\author[d]{Eric Rivals}
\author[a]{Fr\'ed\'eric Geniet}
\author[a]{Nils-Ole Walliser}
\author[a]{John Palmeri}
\author[a]{Andrea Parmeggiani}
\author[a]{Jean-Charles Walter}
\affil[a]{Laboratoire Charles Coulomb (L2C), Univ. Montpellier, CNRS, Montpellier, France}
\affil[b]{Department of Physics, Faculty of Natural and Applied Sciences, Umaru Musa Yar'adua University, Katsina, Nigeria}
\affil[c]{Institut de G\'en\'etique Fonctionelle (IGF), CNRS, Montpellier University, Montpellier, France}
\affil[d]{Laboratoire d'Informatique, de Robotique et de Micro\'electronique de Montpellier (LIRMM), CNRS, Montpellier University, Montpellier, France}
\correspondingauthor{\textsuperscript{2}To whom correspondence should be addressed. E-mail: andrea.parmeggiani@umontpellier.fr,\quad jean-charles.walter@umontpellier.fr}

\date{\today}

\begin{abstract}
Gene expression consists in the synthesis of proteins from the information encoded on DNA. One of the two main steps of gene expression is the translation of messenger RNA (mRNA) into polypeptide sequences of amino acids. Here, by taking into account mRNA degradation, we model the motion of ribosomes along mRNA with a ballistic model where particles advance along a filament without excluded volume interactions. Unidirectional models of transport have previously been used to fit the average density of ribosomes obtained by the experimental ribo-sequencing (Ribo-seq) technique. In this case an inverse fit gives access to the kinetic rates: the position-dependent speeds and the entry rate of ribosomes onto mRNA. The degradation rate is not, however,  accounted for and experimental data from different experiments are needed to have enough parameters for the fit. Here, we propose an entirely novel experimental setup and theoretical framework consisting in splitting the mRNAs into categories depending on the number of ribosomes from one to four. We solve analytically the ballistic model for a fixed number of ribosomes per mRNA, study the different regimes of degradation, and  propose a criteria for the quality of the inverse fit. The proposed method provides a high sensitivity to the mRNA degradation rate. The additional equations coming from using the monosome (single ribosome) and polysome (arbitrary number) ribo-seq profiles enable us to determine all the kinetic rates in terms of the experimentally accessible mRNA degradation rate.
\end{abstract}

\maketitle

\section{Introduction}

The translation of messenger RNA (mRNA) is, together with DNA transcription, one of the two main steps of gene expression. The information encoded into mRNAs is translated into a polypeptidic chain through the action of ribosomes processing along mRNA. Ribosomes match a triplet of nucleic acids (a codon) to the corresponding amino acid. Different kinetic rates are involved during translation: the initiation rate, i.e. the rate at which a ribosome assembles at the start codon of mRNA and starts the translation, the termination rate, i.e. the rate at which the ribosome releases the protein and disassembles from mRNA when it achieves translation, and the hopping rates from codon to codon (elongation rates) which depend on the translated codon (and thus on the position along the transcript) \cite{alberts_molecular_2015}.

Despite its importance in gene expression, and more generally in the metabolism of the cell, much remains to be learned about the kinetics of translation. The onset of the high-throughput technique known as ribosome-sequencing (Ribo-seq) gives access nowadays to genome-wide ribosome density profiles along mRNA \cite{duc_impact_2018}. Ribo-seq is an experimental technique based on the sequencing of ribosome-protected fragments of mRNA after fixation of ribosomes and degradation of unprotected mRNA. Although this approach has revolutionized the study of translation,  one still needs a deeper physical understanding of the molecular processes in order to turn Ribo-seq data into quantitative modeling tools. Ribo-seq data could then be used to probe the microscopic parameters tuning gene expression at the translational level and to understand the role of deregulation of translation in diseases like cancer \cite{brar_ribosome_2015}.

Several approaches have been proposed to describe quantitatively Ribo-seq data by means of physical modeling involving one dimensional transport \cite{valleriani_turnover_2010,siwiak_comprehensive_2010,deneke_effect_2013,duc_impact_2018,fernandes_gene_2017,ciandrini_ribosome_2013,szavits-nossan_inferring_2020}. However, fitting the kinetic model parameters from Ribo-seq data currently faces several major limitations. Firstly, estimating kinetic rates from ribosome profiles is hindered by the lack of sufficient independent experimental data. Since this is a typical inverse problem \cite{tarantola_inverse_2004}, parameters can be under/over-determined creating technical difficulties for making reliable fits. Secondly, most quantitative models assume that mRNAs have infinite lifetimes. The question therefore arises concerning the possible influence of finite mRNA lifetime on the evaluation of ribosome kinetic rates \cite{valleriani_turnover_2010,siwiak_comprehensive_2010,deneke_effect_2013}.

In this article, we model the motion of ribosomes along mRNA using a ballistic model first introduced by Valleriani et al. \cite{valleriani_turnover_2010} of unidirectional transport along a filament including finite mRNA lifetimes. We go beyond  previous approaches, however, by modeling a new experimental procedure where the mRNA population is sorted out into $k$-somes having a fixed number $k$ of ribosomes, going from 1 to 4, present at the instant of the sequencing. Our goal is to demonstrate  how to exploit this experimental data to establish a reliable method for an absolute evaluation of kinetic parameters (i.e., in time units).

Heyer et al. \cite{heyer_redefining_2016} previously presented an interesting pioneering  Ribo-seq study on \textit{Saccharomyces cerevisiae} that was restricted to monosomes and polysomes. According to these authors mRNA translation could in general be divided into three classes. Although they discuss the influence of mRNA lifetime and conclude that their data reveal a relationship between it and the monosome occupancy, they do not propose a mechanistic explanation, but simply note that genes enriched in monosomes had a lower median mRNA half-life than those without enrichment. Despite this finding they do not attempt to integrate mRNA lifetime into their classification. We present here a more quantitative analysis that reveals the crucial importance of mRNA lifetime in classifying genes. Our main result is that although mRNA lifetime does not affect the polysome footprint, it does have a strong impact on the monosome one, despite the low probability of having a monosome.

The paper is organized as follows. In the second section, we define the ballistic model coupled to an mRNA degradation process. In the third section, we solve it analytically by calculating the exact $k$-some probability distribution and density profiles as functions of the kinetic parameters. In the fourth section, we  perform a parametric analysis of these quantities showing that: 1) the smaller the value of $k$, the more these quantities are sensitive to mRNA degradation; 2) for typical biological values of the parameters, low order $k-$somes are more sensitive to mRNA degradation than the full population of mRNAs (polysomes). In the fifth section, we use the model to analyse real Ribo-seq data obtained from a human histonic gene.  For this particular gene, we give - for the first time to our knowledge - evidence of degradation on Ribo-seq data. These results validate our approach and provide motivation for setting up and testing a quantitative method in the sixth section. That method rests on combining $k$-some (monosome actually) data with polysome data while taking advantage of the sharply increased sensitivity of mRNA degradation on low order $k$-somes with respect to polysomes. As a proof-of-concept, we first apply our fitting method to data generated from the ballistic model itself. By using the experimentally accessible degradation rate to fix the time scale, our approach leads to an improved method for obtaining absolute kinetic rates of ribosomes. In the seventh section, we rationalize our findings by proposing a global picture of the different dynamical regimes in the form of a \textit{phase diagram} based on the ballistic model. It presents a condensed global view of the different mRNA degradation dependent regimes and allows at a glance a general qualitative analysis for any given gene.

\section{Definition of the ballistic model and k-somes}

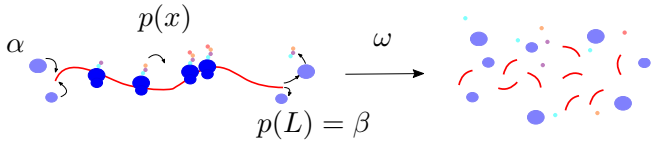
\begin{figure}[h]
    \includegraphics[width=1\columnwidth]{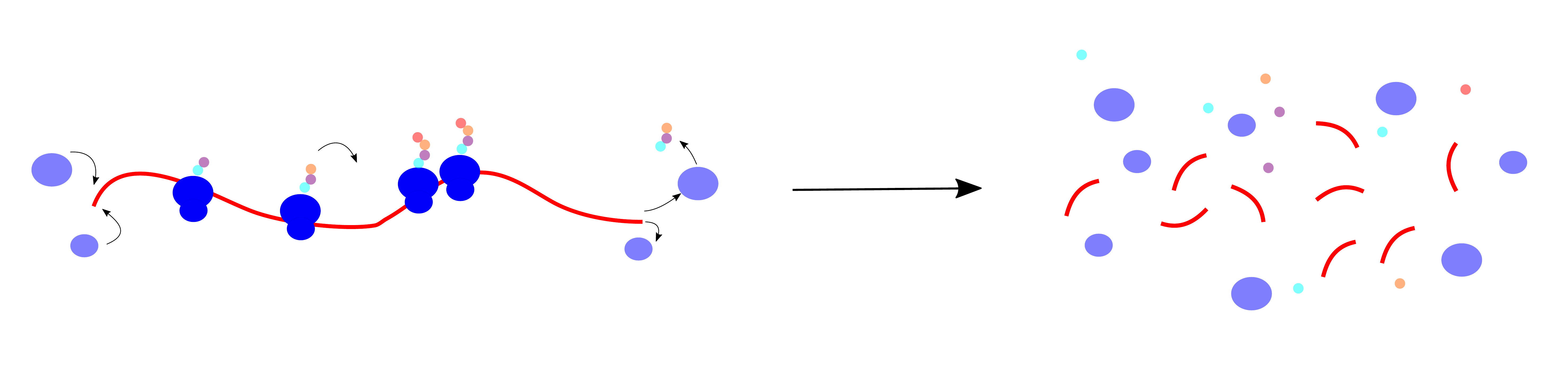} 
     \put(-250,40){\Large $\alpha$}
     \put(-200,50){\Large $p(x)$}
     \put(-111,42.5){\Large $\omega$}
     \put(-155,10){\Large $p(L)=\beta$}
    \caption{Graphical illustration of ribosome transport with the ballistic model including degradation. The mRNA is represented as a red filament of length $L$. The ribosomes are point-like particles processing along the filament with a position dependent rate $p(x)$ (where $x\in[0;L]$ is the position along the filament). Bound and unbound ribosomes are respectively represented by dark blue and light blue discs. A ribosome can bind at the entry of the filament at a rate $\alpha$. A ribosome located at the exit unbinds at a rate $p(L)=\beta$. The mRNAs get degraded at a rate $\omega$.}
    \label{fig:def}
\end{figure}

We model the translation of mRNA by ribosomes, and the degradation of mRNA, by means of a ballistic model with degradation as sketched in Figure~\ref{fig:def}. The mRNA is modeled by a filament of length $L$ where ribosomes are treated as point-like particles. i.e. without excluded volume interactions. This is motivated by the low density of ribosomes on mRNA. Particles get attached at the entrance of the filament with a stochastic rate $\alpha$ (initiation rate at the start codon) and subsequently progress in one direction along the filament with a local deterministic velocity $p(x)$ (related to the local elongation rate), that depends on the position $x$ along the transcript. We shall simply use $p(x)=p$ for a constant (codon independent) hopping rate. Particles leave the filament with velocity $p(L)$ (termination rate at the stop codon). 
We emphasize that our model does not account for excluded volume interactions, thus traffic jam effects induced in the Totally Asymmetric Simple Exclusion Process - TASEP \cite{macdonald1968kinetic}, the paradigmatic model for translation modeling, are not possible. The ballistic model, however, remains a good approximation to TASEP for long enough mRNA lifetimes in the low density regime, which is the relevant regime for translation \cite{yahaya-in-preparation}.

Let $\mathcal T(x)$ be the time for a ribosome to go from genomic coordinate 0 to $x$. Since the ballistic trajectory of a ribosome, $X_0(t)$, obeys $dX_0(t)/dt = p(x)$, we obtain
\begin{equation} \label{Tx}
\mathcal{T}(x)=\int_0^x\frac{dy}{p(y)}\,.
\end{equation}
The particular value $\mathcal{T}(L)$ corresponds to the time needed for a ribosome to cross the entire mRNA. This quantity  plays an important role in the determination of the average density profiles of ribosomes along mRNA.

Our model includes the degradation of mRNA at a constant rate $\omega$ as in Ref.~\cite{valleriani_turnover_2010}. 
For the sake of simplicity, we assume instantaneous  mRNA degradation -- translation is therefore instantaneously aborted -- but other, more progressive, mechanisms exist~\cite{valleriani_turnover_2010,deneke_effect_2013}. We assume that mRNAs are synthesized and degraded in the cytosol in such a way that their population is constant on average. In what follows, we shall refer to the "age", $a$, of an mRNA as the time elapsed between its synthesis and the time at which the Ribo-seq experiment is actually done. The mRNA lifetime $\theta$ is modeled by a random variable with probability density function $\phi_{\theta}(\theta)$. The assumption of stationarity of the mRNA population leads to a relation between the mRNA age distribution, $\phi_a(a)$, and the lifetime distribution $\phi_{\theta}(\theta)$~\cite{cox_renewal_1970}
\begin{equation}
    \phi_a(a)=\frac{1}{\langle\theta\rangle}\int_a^\infty\phi_\theta(\theta)d\theta\, , 
\end{equation}
where $\langle\theta\rangle=\int_0^\infty\theta\phi_\theta(\theta) d\theta$ is the average lifetime of an mRNA. 
As in \cite{valleriani_turnover_2010}, we shall consider an exponential mRNA lifetime distribution  
$\phi_\theta(\theta)=\omega e^{-\omega \theta}$, $\theta \geq 0$, 
with average lifetime $\langle\theta\rangle=\omega^{-1}$. The mRNA age distribution is thus also exponential and can be written as
\begin{equation} \label{Phiaexp}
    \phi_a(a)=\omega e^{-\omega a}\,, \ a \geq 0\, .
\end{equation}

In a biological cell, mRNA can be found in different states: newly synthesized mRNA empty of ribosomes, mRNA in a transient state evolving from an empty to stationary state, and finally mRNA in the stationary state of translation. We divide the whole population of mRNA into sub-populations, denoted $k$-somes containing exactly  $k$ ribosomes. The first four categories containing from one to four ribosomes, respectively, are called monosomes, disomes, trisomes and tetrasomes. 

We will show in the next sections that these $k$-somes are more sensitive to mRNA degradation than the whole population of intracellular mRNAs (polysomes) and that these independent density profiles provide sufficient information for estimating all relative translation kinetic rates in a single experiment.

\section{Theoretical results}

In this section we present an analysis of the $k-$somes within the framework of the ballistic model.

\subsection{Distribution of the $k$-some population} 

The number of ribosomes on a particular mRNA depends on its age $a$: if $a < \mathcal{T}(L)$, ribosomes present on the mRNA at the time of the experiment have been loaded during the time interval $a$ preceding the experiment while for $a > \mathcal{T}(L)$, they have been loaded during the time interval $\mathcal{T}(L)$ before the experiment. The loading process being of the Poisson type, the probability that an mRNA of age $a$ be loaded with $k$ ribosomes is therefore given by 
\begin{equation} \label{Pka}
 P(k|a) = \begin{cases}
\frac{(\alpha a)^k}{k!} e^{-\alpha a},\ a \leq {\cal T}(L)\,,\\
\frac{(\alpha {\cal T}(L))^k}{k!} e^{-\alpha {\cal T}(L)},\ a \geq {\cal T}(L)\,.
\end{cases}
\end{equation}
Using \eqref{Phiaexp}, the probability $P_k$ that an mRNA be a $k$-some irrespective of its age is then given by (see SI, section 3.A, for an explicit analytical expression)
\begin{equation} \label{Pk1}
P_k = \int_0^{\infty}\!\!  P(k|a) \phi_a(a)\, da    \, .
\end{equation}
 
\subsection{$k$-some density profile along mRNA}

Another important result about Poisson processes is that, knowing that $k$ events have occurred during a given time interval, the probability density of their occurrence times is uniform on that time interval (see for example Ref. \cite{d_j_daley_introduction_2003} for a rigorous statement of that theorem). Given the one-to-one correspondence between the ribosome initiation time and its location at the time of the Ribo-seq experiment, using the ribosome equation of motion, $d\mathcal T(x)/dx = 1/p(x)$, the density profile $\rho_k(x|a)$ of a $k$-some knowing that its age is $a$ can be written as
\begin{equation} \label{rhokxa}
\rho_k(x|a) = 
\begin{cases} 
\frac{k}{a p(x)} H(x(a)-x), \ a \leq {\cal T}(L)\,,\\
\frac{k}{{\cal T}(L) p(x)}, \ a \geq {\cal T}(L)\, ,
\end{cases} 
\end{equation}
where $H(x)$ stands for the Heaviside function. In the first line of \eqref{rhokxa}, $x(a)$ is the maximal distance that a ribosome can possibly travel along an mRNA with an age $a<\mathcal T(L)$. It is given implicitly by
$a = \int_0^{x(a)} dy/p(y)$.
\eqref{rhokxa} is easily interpreted: for recently synthesized mRNA such that $a < \mathcal T(L)$, a ribosome reaches at most the distance $x(a)$ along the mRNA. The density is therefore zero beyond that limit (hence the Heaviside function) and the system behaves as if the mRNA had an effective length  $x(a) < L$.
Moreover, the ribosome density is proportional to the time ribosomes spend at location $x$ which is inversely proportional to their local velocity $p(x)$. 

According to \eqref{rhokxa}, it is possible to distinguish two states for mRNA with translating ribosomes: a {\it transient} state, for which young mRNA with $a < \mathcal T(L)$ have densities $\rho_k(x|a)$ that depend  explicitly  on the age $a$, and a {\it stationary} state for which $a > \mathcal T(L)$ and $\rho_k(x|a)$ no longer depends explicitly on $a$. We will see later that the proportion of mRNA in one state versus the other is critical in characterizing the age-averaged density.
Integrating any of these densities over the entire mRNA length yields, as expected, the number of ribosomes, $k$.

The density profile of a $k$-some {\it irrespective} of mRNA age - which is experimentally accessible for small values of $k$ - is given by
\begin{equation} \label{rhokx1}
\rho_k(x) = \int_{0}^{\infty}  \!\! \rho_k(x|a) p(a|k) \, da,
\end{equation}
where $p(a|k)$ is the age distribution among $k$-somes given by Bayes theorem, $p(a|k) = \phi_a(a) P(k|a)/P_k$. The analytical expression for $\rho_k(x)$ is somewhat complex and shall be provided and analysed in section \ref{sec:ksome_density_analysis}.

\subsection{Polysome density profile along mRNA}

The ribosome density profile obtained for the entire mRNA population (polysomes) may be calculated
as the average over $k$ of the $k$-some density profiles
\begin{equation}
 \rho(x)= \sum_{k=0}^{\infty} \rho_k(x) P_k =   \frac{\alpha}{p(x)}e^{-\omega{\mathcal T}(x)}\,.
 \label{rho_vs_p}
\end{equation}
Integrating over $x$ leads to the average number of ribosomes $\langle k\rangle$ loaded on an mRNA
\begin{equation}
    \langle k \rangle =\frac{\alpha}{\omega}\left(1-e^{-\omega\mathcal T(L)} \right)\,.\label{k_av}
\end{equation}
The two former equations, \eqref{rho_vs_p} and \eqref{k_av}, are in agreement with Ref.\cite{valleriani_turnover_2010}, showing the consistency of our approach. Note that the result \eqref{k_av} can also be obtained from $\langle k \rangle = \sum_{k=0}^{\infty} k P_k$. Note also that, in the limit of infinite lifetime ($\omega=0$), the polysome density \eqref{rho_vs_p} becomes $\rho^\infty(x)=\alpha/p(x)$, a result that simply expresses the
 conservation of the local particle current ($j(x) = \rho(x) v(x) = {\rm cst}$) in the stationary  regime, and that can indeed be derived in the TASEP low density phase \cite{Derrida_1993}.

\section{Analysis of finite lifetime effects } 

In this section we rationalize the parametric dependence of the previous expressions in the relevant physical regimes. We shall use ${\cal T}(L)$ as a time scale throughout the rest of the paper. This allows us to  define dimensionless rates (with tildes) as original ones multiplied by ${\cal T}(L)$
\begin{equation}
\tilde{\alpha} = \alpha {\cal T}(L)\ ;\ \tilde{\omega} = \omega {\cal T}(L)\ ;\ 
\tilde{p}(x) = p(x) {\cal T}(L)\, .
\end{equation}

\subsection{Values of kinetic rates in various organisms} 
Since we are studying the effects of  finite mRNA lifetime, we report in table \ref{tab:rates_bio}  some literature values for the  kinetic parameters and in table \ref{tab:dimless_param_bio} their dimensionless counterparts that prove essential in determining typical biological regimes. We note that typical mRNA lifetimes for different species can span two orders of magnitude, while molecular  quantities vary significantly less (e.g. a factor 7 for $p$, a factor 2 for $L$, a factor 10 for $\mathcal{T}(L)$).   
\begin{center}
  \begin{table}[h]
   \begin{tabular}{|*{6}{c|}}
     \hline
     Species & \makecell{mRNA $t_{1/2}$ \\(median)} & \makecell{ $p$ \\$(s^{-1})$ }& \makecell{ $L$\\(codons)} & \makecell{$\alpha$ \\ $(s^{-1})$} & \makecell{$\mathcal{T}(L)$\\$(s)$} \\ \hline \hline
    {\it E. coli} & 5 min  & 10-20 & 256& -  & 13-26 \\ \hline
    {\it S. cerevisiae} & 20 min & 3-10 & 399 & 0.1 & 40-133 \\ \hline
    {\it M. musculus} & 7 h  & 6 & 493 & - & 82\\ \hline
    {\it H. sapiens} & 9 h & 3 & 481 & 0.06 & 160 \\ \hline
    \end{tabular}
    \caption{Translation kinetic rates and mRNA properties in bacteria, yeast and mammals. Medians of gene length $L$ are calculated from the entire genomes set available on the NCBI database. mRNA half-lives $t_{1/2}$ \cite{bernstein_global_2002,wang_precision_2002,sharova_database_2009,schwanhausser_global_2011}, elongation rates $p$ \cite{young_polypeptide-chain-elongation_1976,karpinets_rnaprotein_2006,ingolia_ribosome_2011,yan_dynamics_2016} and initiation rates $\alpha$ \cite{ciandrini_ribosome_2013,trosemeier_optimizing_2019} are taken from the literature. $\mathcal{T}(L)$ is approximated by $L/p$ (unit length is expressed in codons).        
    }
    \label{tab:rates_bio}
   \end{table}
 \end{center}
 
 \begin{center}
   \begin{table}[h]
   \begin{tabular}{|*{5}{c|}}
     \hline
     Species & $\tilde{\omega} = \omega \mathcal{T}(L)$ & $\tilde{\alpha} =\alpha\mathcal{T}(L)$& $\omega/\alpha$&$\mathcal R_1(0)$ \\ \hline \hline
    {\it E. coli}  & 0.04-0.07 &-&-& -\\ \hline
    {\it S. cerevisiae}  & 0.02-0.08 & 4-13& 6$\times10^{-3}$ & 0.3-3700 \\ \hline
    {\it M. musculus}  & 0.002 &-&-&- \\ \hline 
    {\it H. sapiens}  & 0.003 & 10 & 3$\times10^{-4}$ & 5  \\ \hline
   \end{tabular}
   \caption{Calculated dimensionless parameters in bacteria, yeast and mammals from parameters taken from TABLE \ref{tab:rates_bio}. The quantity $\mathcal R_1(0)$ will be defined in the next section, see Eq.(\ref{ratio1}).}
   \label{tab:dimless_param_bio}
  \end{table}
 \end{center}
\subsection{Polysomes are not sensitive to mRNA degradation} 
\label{secB}
\begin{figure}[h]
    \centering\includegraphics[width=0.85\columnwidth]{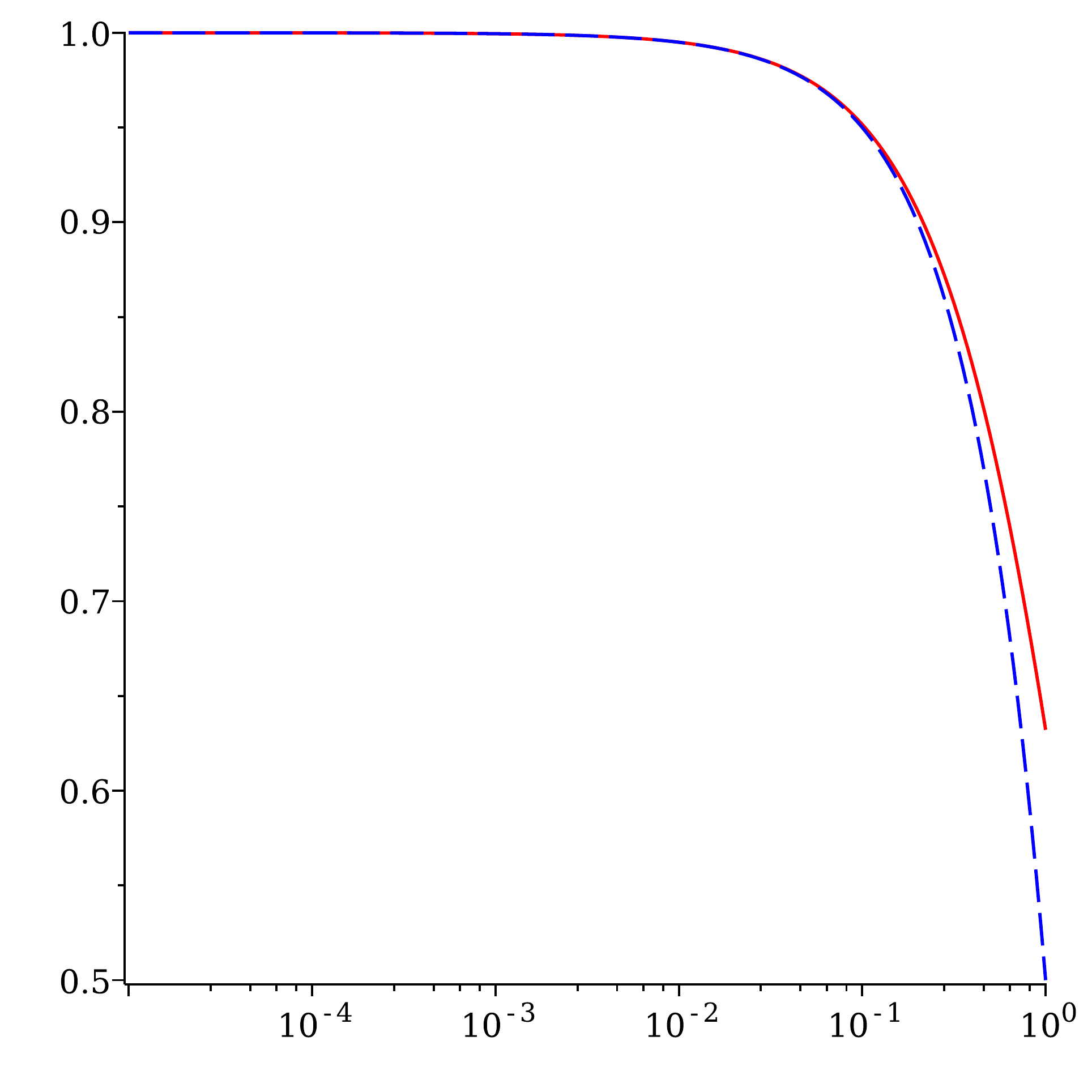} 
    \put(-210,100){\small {\rotatebox{90}{\Large $f(\tilde\omega)$ }} }
    \put(-100,-7){\Large $\tilde\omega$}
    \vspace{0.2cm}
    \caption{Horizontal semilog plot of the function  $f(\tilde\omega)=\langle k \rangle / \tilde\alpha$ defined in Eq.(\ref{fxi}) vs $\tilde\omega=\omega\mathcal T(L)$ (solid red line) and asymptotic behavior when $\tilde\omega\to0$, $f(\tilde\omega) \simeq 1-\frac{\tilde\omega}{2}$ (dashed blue line). The function has been plotted over 5 biologically relevant decades: $\tilde\omega \in [10^{-5},1]$.} 
    \label{fig:k_av}
\end{figure}

In this subsection, we discuss the effect of degradation on the whole population of mRNA. 
In doing so, we shall define $f(\tilde\omega)$ as the average number of ribosomes $\langle k \rangle$ normalized by the same quantity in the infinite lifetime limit, $\langle k \rangle_\infty=\tilde{\alpha}$ (see Eq.(\ref{k_av})):
\begin{equation}
f(\tilde\omega)=\frac{\langle k \rangle}{\langle k \rangle_\infty}=\frac{\langle k \rangle}{\tilde\alpha}=\frac{1}{\tilde\omega}\left(1-e^{-\tilde\omega}\right)\,.
\label{fxi}
\end{equation}
This function can be interpreted in terms of a birth and death process (see SI, section 2).
It only depends on $\tilde\omega$ and is plotted in Figure \ref{fig:k_av} in semilog scale. 

As expected, $f(\tilde\omega)$ decreases with increasing $\tilde\omega$: the average number of ribosomes per mRNA decreases as the mRNA lifetime becomes shorter. Nonetheless,
biological values from the literature presented in Tab. \ref{tab:rates_bio}  and \ref{tab:dimless_param_bio} show that in yeast and bacteria, $\tilde{\omega}\sim 10^{-2}$, while in mammals $\tilde{\omega}\sim 10^{-3}$. Therefore, $\tilde\omega\ll1$ for the majority of genes and $f(\tilde\omega)\sim 1 - \tilde\omega/2$ remains very close to unity, showing that degradation has very little impact on $\langle k \rangle$, the most relevant parameter characterizing polysomes. 

To strengthen even more this argument, let us discuss the polysome density profile, $\rho(x)$, which is one of the  experimentally accessible quantities measured in Ribo-seq experiments, and compare it to its infinite lifetime limit,
$\rho^\infty(x)=\alpha/p(x)$. We see from Eq. (\ref{rho_vs_p}) that the influence of mRNA degradation on $\rho(x)$ is maximal at the last site (i.e., $x=L$), which leads to the following bound
\begin{equation} \label{rho_relat_err}
 \forall x \in [0,L],\quad   \frac{|\rho(x)-\rho^\infty(x)|}{\rho^\infty(x)} \leq 1-e^{-\tilde\omega}  \approx \tilde\omega,
\end{equation}
where the approximation holds for $\tilde\omega \ll 1$. For $\tilde\omega = 10^{-1}$, an upper bound of the values indicated in table \ref{tab:dimless_param_bio}, replacing the polysome density by its infinite lifetime limit would lead to a relative error of no more than 10\%. Therefore, no modifications due to mRNA degradation should be detectable in polysome Ribo-seq experiments. This analysis shows that previous works~\cite{ciandrini_ribosome_2013, duc_impact_2018} estimating the hopping rates in {\it S. cerevisiae} from polysomes with a model that does not take into account the mRNA finite lifetime provide a valid approximation. At variance, we shall demonstrate in the next sections that, in the same conditions, $k$-somes are more sensitive to degradation than polysomes and constitute a key tool to determining kinetic rates. 

\subsection{$k$-some analysis} 
\subsubsection{$k$-some distribution} 

\begin{figure}[h]
    \centering
    \includegraphics[width=1\columnwidth]{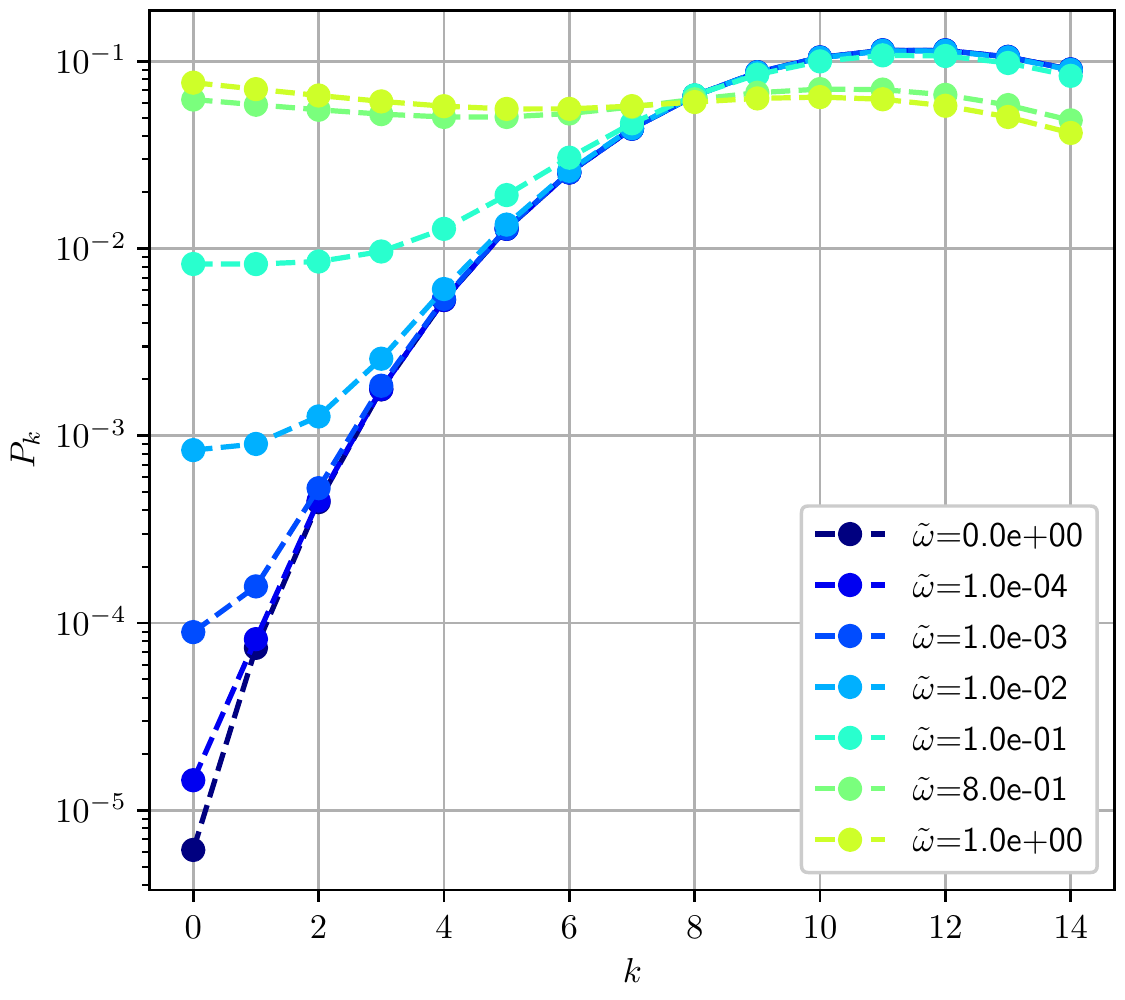} 
        \begin{picture}(0,0)
        \put(5,125){\includegraphics[height=2.3cm]{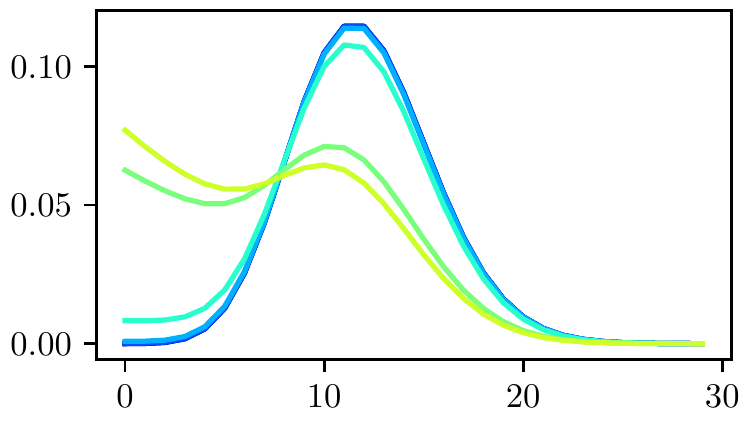}}
        \end{picture}
    \caption{The $k$-some population distribution $P_k$ for finite values of $\tilde\omega$ (from $0$ to 1) and other parameters fixed: $\mathcal T(L)=200s$, $L=100$ codons and $\alpha=0.06s^{-1}$. The value $\langle k\rangle=\tilde\alpha=12$ corresponds to the maximum of the distribution at $\tilde\omega=0$. In the main graph (lin-log scale) we can appreciate the increase of the population of empty mRNA ($k=0$) and small value of $k$ as $\tilde\omega$ increases. In the inset (lin-lin), we can see that the two maxima (at $k=\tilde\alpha$ and $k=0$) equate for $\tilde\omega=0.8$. 
    Expression (\ref{Pk_linear_omega}) for $P_k$  provides an excellent approximation for  $k \le 4$, except for the highest two values of $\tilde\omega$ (for which the approximation remains good with less than 20\% error for $k \le 1$).}
    \label{fig:theory2}
\end{figure}
In terms of the dimensionless parameters $\tilde{\alpha}$ and $\tilde{\omega}$, the probability $P_k$ (see SI, section 3.A) that an mRNA be a $k$-some writes 
\begin{multline}
 P_k = 
 \frac{\tilde{\omega}}{\tilde{\alpha}+\tilde{\omega}}\left(\frac{\tilde{\alpha}}{\tilde{\alpha}+\tilde{\omega}}\right)^k\frac{\gamma(k+1,\tilde{\alpha}+\tilde{\omega})}{k!} +\frac{\tilde{\alpha}^k}{k!}e^{-(\tilde{\alpha}+\tilde{\omega})}\,,
 \label{Pkn}
\end{multline} 
where details about the incomplete Gamma function, $\gamma(k,z)$, may be found in the SI, section 1.
In Figure~\ref{fig:theory2}, this distribution is plotted for typical values of the parameters: $\mathcal T(L)=200s$,  $L=100$ codons, $\alpha=0.06s^{-1}$ ($\tilde\alpha=12$) and different values of $\tilde\omega$ between 0 (no degradation) and 1 (high degradation). A log-lin scale is used in the main graph to better observe the evolution of the $k$-some population with small $k$. As $\tilde\omega$ increases, we see that the populations of empty mRNA (zerosomes) and of the mono- through tetrasome increase quasi linearly with $\tilde\omega$ and that the distribution progressively displays a plateau for low values of $k$. This result is best understood by carrying out a linear expansion of $P_k$ in $\tilde\omega$ for $\tilde\alpha \gg k$:
\begin{equation} \label{Pk_linear_omega}
    P_k \simeq P_k^{\infty} + \frac{\tilde\omega}{\tilde\alpha},\quad \textrm{with}\quad P_k^{\infty}=\frac{\tilde\alpha^k e^{-\tilde\alpha}}{k!}\,.
\end{equation}
For values of $\tilde\alpha \sim 10$, the first term of this expansion, $P_k^{\infty}$ - that is of the Poisson type, as expected given the nature of the ribosome loading process - remains very small compared to the linear term in $\tilde\omega$. We therefore have $P_k \sim \tilde\omega/\tilde\alpha$, a result both linear in $\tilde\omega$ and independent of $k$ that explains the observed plateau. For $\tilde\omega < 10^{-1}$, when $k$ increases, however, the $k$-some distribution becomes poissonian again and $P_k \sim P_k^{\infty}$, as we may observe in Fig. ~\ref{fig:theory2}. 
$ $ 

In the inset, the same data are plotted on a lin-lin scale and with a broader range for $k$. For $\tilde\omega < 10^{-1}$, we observe a nearly symmetric Gaussian distribution that mainly corresponds to the Poisson term $P_k^{\infty}$. It is centered around a maximum close to $k = \langle k\rangle_{\infty}=\tilde\alpha=12$, the value of the average number of ribosomes in the infinite lifetime limit. As expected, most mRNAs are then loaded with that number of ribosomes when $ \tilde\omega < 10^{-1}$. For higher values of $\tilde{\omega}$, however, typically when $\tilde{\omega} > \tilde\omega_{\rm inv}$ where $\tilde\omega_{\rm inv}=\ln \left( \sqrt{\frac{\tilde{\alpha}}{2 \pi}} + \frac{1}{2} \right)
$ (see SI, section 3.D), an inversion of the populations occurs: the proportion of small order $k$-somes takes over the population of those for which $k\sim \tilde\alpha$ (local maximum). This result corresponds to a regime where mRNAs do not live long enough for ribosomes to reach the exit, hence reducing their number and favoring $k$-some populations with small $k$. For biologically relevant values ($\tilde\alpha \sim 10$),  $\tilde\omega_{\rm inv}$ is on the order of unity, in good agreement with the value $\tilde\omega \sim 1$ for which the polysome density is significantly impacted by degradation (see \eqref{rho_relat_err}).

\subsubsection{$k$-some densities} %
\label{sec:ksome_density_analysis}

One might suspect from the previous results that, due to their statistical prominence when $\tilde\omega \sim 1$, mono- through tetrasomes ($k$-somes with small $k$) might play a decisive role in the determination of the kinetic parameters only once that condition is fulfilled. This is however not true for, as we shall see, $k$-some densities become 
sensitive to mRNA degradation for smaller values of $\tilde\omega$ than those ensuring their dominance.
Therefore, inasmuch as a statistically relevant sample of them (say thousands) is available in Ribo-seq
data, densities may be analysed to determine the kinetic parameters. Hundreds of millions of mRNA sequences 
are typically involved in Ribo-seq experiments. This means that for 
$\tilde\alpha \sim 10$ and as soon as $\tilde\omega > 10^{-4}$, thousands of monosomes and even more of $k$-somes with $k>1$ are available for analysis, enough to reach a good statistical accuracy. 

In terms of dimensionless parameters, the exact $k$-some density \eqref{rhokx1} reads
\begin{multline} \label{rhokx}
\rho_k(x) =\frac{\tilde\alpha^k}{P_k\,\tilde{p}(x)}\frac{e^{-(\tilde\alpha+\tilde\omega)}}{(k-1)!} +\\
\frac{\tilde\omega}{P_k \tilde{p}(x)}\left(\frac{\tilde\alpha}{\tilde\alpha+\tilde\omega}\right)^k\frac{\gamma(k,\tilde\alpha+\tilde\omega)-\gamma(k,(\tilde\alpha+\tilde\omega)\tau(x))}{(k-1)!}\,,
\end{multline}
where $\tau(x)=\mathcal T(x)/\mathcal T(L) \in [0,1]$. This expression is made of two terms, each corresponding to those obtained in \eqref{rhokxa}. The first one represents the contribution to the $k$-some density of mRNAs that have reached their stationary state of translation in the sense that a ribosome may have had the chance to exit them prior to the experiment.  We shall refer to that term as $\mathcal{S}_k(x)$. The second term corresponds to mRNAs that are still in their transient state of translation: the time elapsed between their synthesis and 
the Ribo-seq experiment is too short for a ribosome to reach their exit codon. We shall call that term $\mathcal{F}_k(x)$. 

To quantify the influence of mRNA degradation on $k$-somes, we shall define a transient to stationary mRNA density ratio as $\mathcal R_k(x)=\mathcal{F}_k(x)/\mathcal{S}_k(x)$. A mathematical analysis of this ratio is done in the SI, section 5.  Its maximal value, achieved at $x=0$, is given by 
\begin{equation} \label{ratiok}
    \mathcal R_k(0)=\Tilde{\omega}\frac{\gamma(k,\Tilde{\omega}+\Tilde{\alpha})}{   (\Tilde{\omega}+\Tilde{\alpha})^k     }e^{\Tilde{\omega}+\Tilde{\alpha}} \simeq \frac{\tilde\omega}{kP_k^{\infty}}\, ,
\end{equation}
where $P_k^{\infty}$ is given in \eqref{Pk_linear_omega}. The last approximation is valid for small values of $\tilde\omega$ and $\tilde\alpha \gg k$ but the exact expression, valid for all values of $\tilde\alpha$
and $\tilde\omega$, allows us to analyse the impact of degradation in the entire parameter space. The fact that
$\mathcal R_k(x)$ achieves a maximal value at $x=0$ means that $k$-some densities are more
sensitive to degradation near the entrance site than towards the end, at variance with polysomes for which it is 
the opposite (see section \ref{secB}). We shall see below that the approximate value of $\mathcal R_k(0)$ appears as a natural parameter in the comparison of the $k$-some density to its infinite lifetime limit. 

\begin{figure}[h!]
    \centering
        \includegraphics[width=0.9\columnwidth]{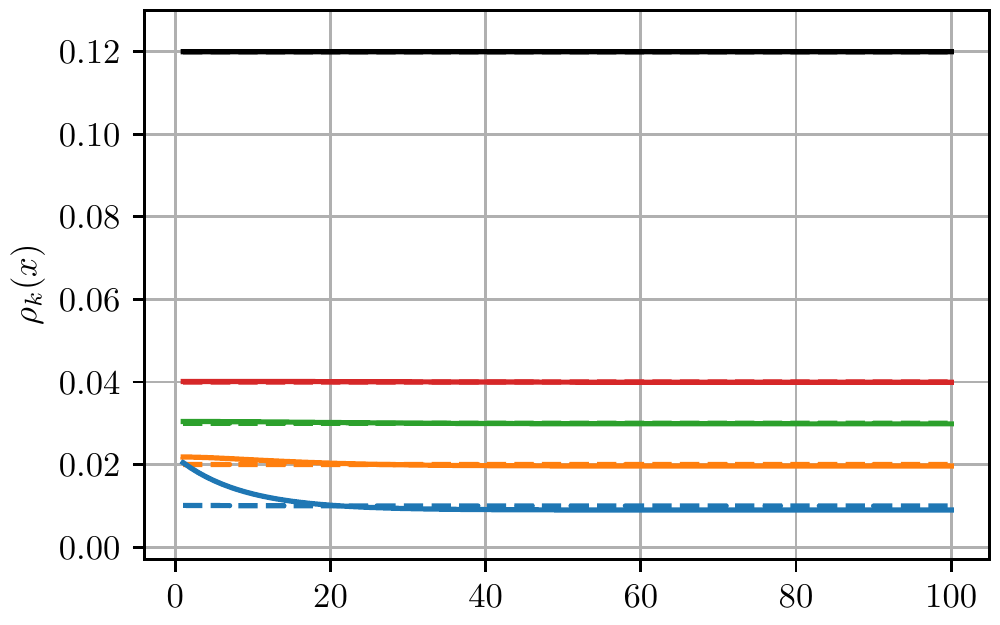}       
        \includegraphics[width=0.9\columnwidth]{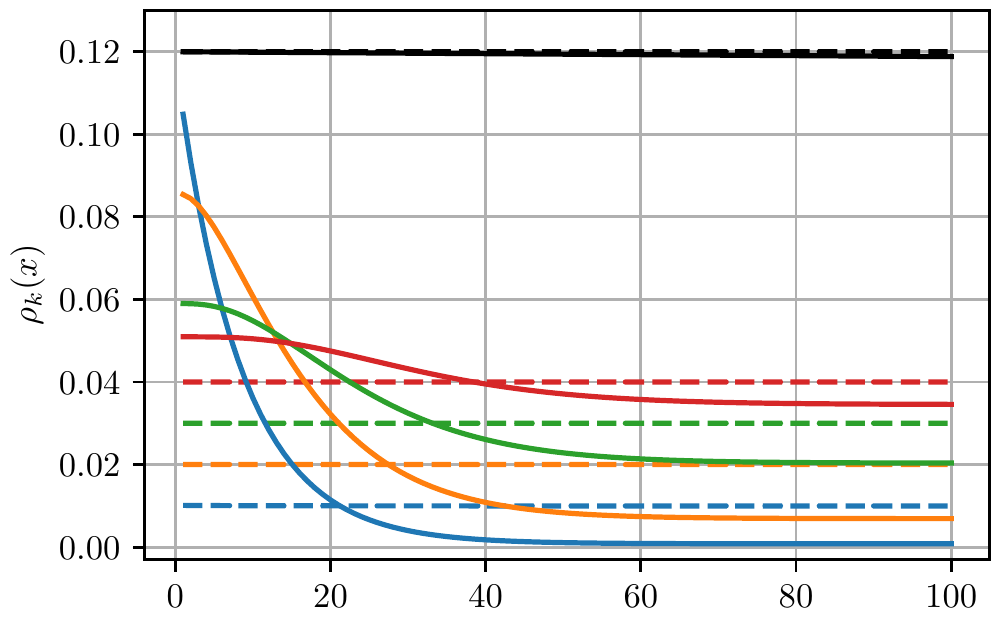}
        \includegraphics[width=0.9\columnwidth]{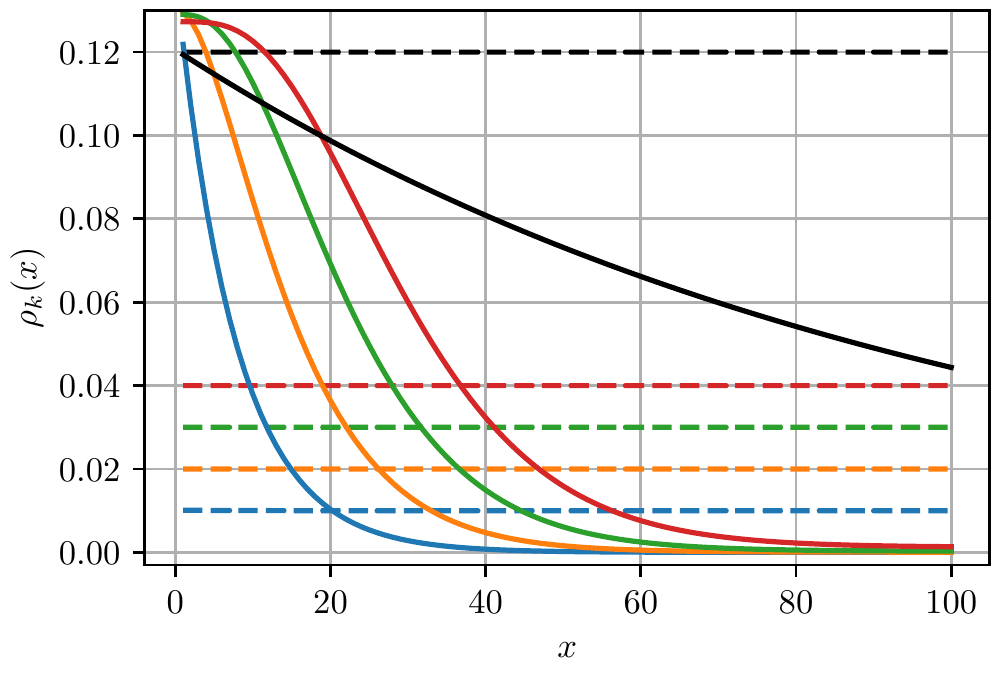}
    \caption{Plot of the $k$-some densities $\rho_k$ for $k=1$ to 4 versus the genomic coordinate $x$ obtained from \eqref{rhokx}: monosome (blue), disome (orange), trisome (green) and tetrasome (red). Black curves correspond to polysome densities. Parameters are $\alpha=0.06s^{-1}$, $p=1/2s^{-1}$ and $L=100$ codons ($\tilde\alpha=12$).  Solid lines are obtained from \eqref{rhokx} for increasing finite values of $\omega=5.10^{-7}s^{-1}$ ($\tilde\omega=10^{-4}$), $\omega=5.10^{-5}s^{-1}$ ($\tilde\omega=10^{-2}$) and $\omega=5.10^{-3}s^{-1}$ ($\tilde\omega=1$) from top to bottom; while dashed lines are obtained for $\omega=0s^{-1}$ (infinite mRNA lifetime).
    }
    \label{fig:theory1}
\end{figure}

To illustrate the global trends of $k$-some densities without further complications, we shall now assume a constant (codon independent) elongation rate\footnote{Note that, a non uniform elongation rate would just modulate the $k$-some density without affecting its global behavior.}, $\tilde{p}(x)=\tilde{p}$.  In that case, the dimensionless time is simply $\tau(x)=x/L$ and the sole term of \eqref{rhokx} that depends on $x$ is $\mathcal{F}_k(x)$. This term is responsible for the global exponential decay of the densities observed in Figure~\ref{fig:theory1} where $\rho_k(x)$ has been plotted for $k=1$ to 4. In that figure, density profiles in solid lines are given by \eqref{rhokx} (the discrete version of which can be found in the SI, section 6) with $L=$100 codons, $\alpha=0.06s^{-1}$, $p=0.5s^{-1}$ and thus $\mathcal T(L)=200s$. Dimensionless parameters are thus $\tilde\alpha=12$ and $\tilde{p}=100$.

In each graph, dashed lines represent the density profiles for an infinite mRNA lifetime ($\tilde\omega=0$) 
which are flat with magnitude $\rho_k^{\infty}=k/L$. In that same limit, the polysome density is also flat, 
with magnitude $\rho^{\infty}=\alpha/p$.  As an important consequence, when $\tilde\omega=0$, all $k$-some and the polysome profiles are linearly related. We shall
see in section \ref{section_comparison} that this precludes the determination of all kinetic parameters in that limit. 

From top to bottom, the values of $\tilde\omega$ used in the successive panels are spaced by a factor of two orders of magnitude: $\tilde\omega=10^{-4}$, $\tilde\omega=10^{-2}$ and $\tilde\omega=1$ corresponding to 555h, 5h and 3 min lifetimes, respectively.
Below, we briefly describe the main features of the densities $\rho_k(x)$. A more detailed analysis is provided in the SI, section 4.

In the first case, all density profiles are close to the $\tilde\omega=0$ limit and we barely notice a small deviation for monosomes, at the start of the filament. Thus, in this regime of low degradation, we do not expect $k$-somes to bring more information than the polysomes in a usual Ribo-seq experiment. 

In the second case, we do not see any effect on the polysome profile, but do see a clear effect on the $k$-somes. Note that, as $k$-some profiles are no longer linearly related, a fit of the kinetic rates is now possible. This decoupling indeed provides extra information for determining the unknown model parameters. Importantly, this intermediate regime  ($\tilde\omega\sim 10^{-2}$) falls into the range of biological values. Additionally, we observe that the ordering of $k$-some densities change with the position along the mRNA: while densities obey $\rho_1>\rho_2>\dots>\rho_4$ close to the initiation site, they switch in reverse order after a few sites.   

In the last case, degradation is high ($\tilde\omega=1$). All density profiles are modified, even the polysomes. This regime is probably not relevant for typical translation processes, but we describe it for the sake completeness and because it might be relevant under extreme cellular conditions.

Let us now turn to the sensitivity of $k$-some density profiles to mRNA degradation. Taking advantage
of the smallness of $\tilde\omega$ compared to other parameters, we carry out a linear expansion of $\rho_k(x)$ in $\tilde\omega$. Yet, as we may observe in Fig.~\ref{fig:theory1}, $k$-some densities are more sensitive to degradation near the ribosome initiation site ($x=0$) than further along the mRNA. More details about the subtle question of the position dependent sensitivity of $k$-some densities are provided in the SI, section 4.B, where the intersection of the $k$-some density profiles with their infinite lifetime limit (flat profiles) is analysed.  Hereafter, we evaluate their sensitivity to $\tilde\omega$ by calculating their relative difference with their infinite lifetime limit at $x=0$.
For $\tilde\alpha \gg k$, we obtain
\begin{equation}
   \frac{\left|\rho_k(0)-\rho_k^{\infty}(0)\right|}{\rho_k^{\infty}(0)} \simeq \frac{\tilde\omega}{kP_k^{\infty}} \simeq \mathcal R_k(0)\, .
\end{equation}
Thus, in the biologically relevant limit of small $\tilde\omega$ values, the transient to stationary mRNA ratio $\mathcal R_k(0)$ controls 
the impact of degradation on $k$-some densities near the initiation site. 

In conclusion, the ballistic model mainly displays three regimes of degradation: (i) low degradation (LD), where $k$-somes and polysomes are linearly related and where $k$-somes are therefore not expected to provide more information than polysomes; (ii) an intermediate regime (ID), into which some genes are likely to fall, where $k$-some densities are already sensitive to degradation and no longer linearly related while polysomes are still unaffected; (iii) a regime of high degradation (HD), not relevant for typical translation processes, in which all profiles are strongly modified by degradation, polysomes included.  

\section{Comparison with experiments} \label{section_comparison}

\subsection{Ribo-seq experiments with $k-$somes}

In this section, we compare the phenomenology of the ballistic model for $k$-somes to experimental data for a gene coding for histones. This comparison will show that the effect of degradation is visible in certain genes and the sensitivity to degradation is enhanced for $k$-somes with respect to polysomes, thus justifying the introduction of this new experimental setup.

\begin{figure}[h!]
    \centering
    \includegraphics[width=0.8\columnwidth]{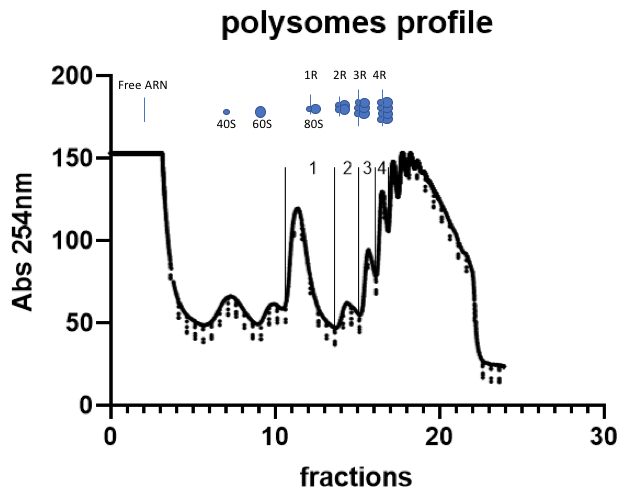}
    \includegraphics[width=0.9\columnwidth]{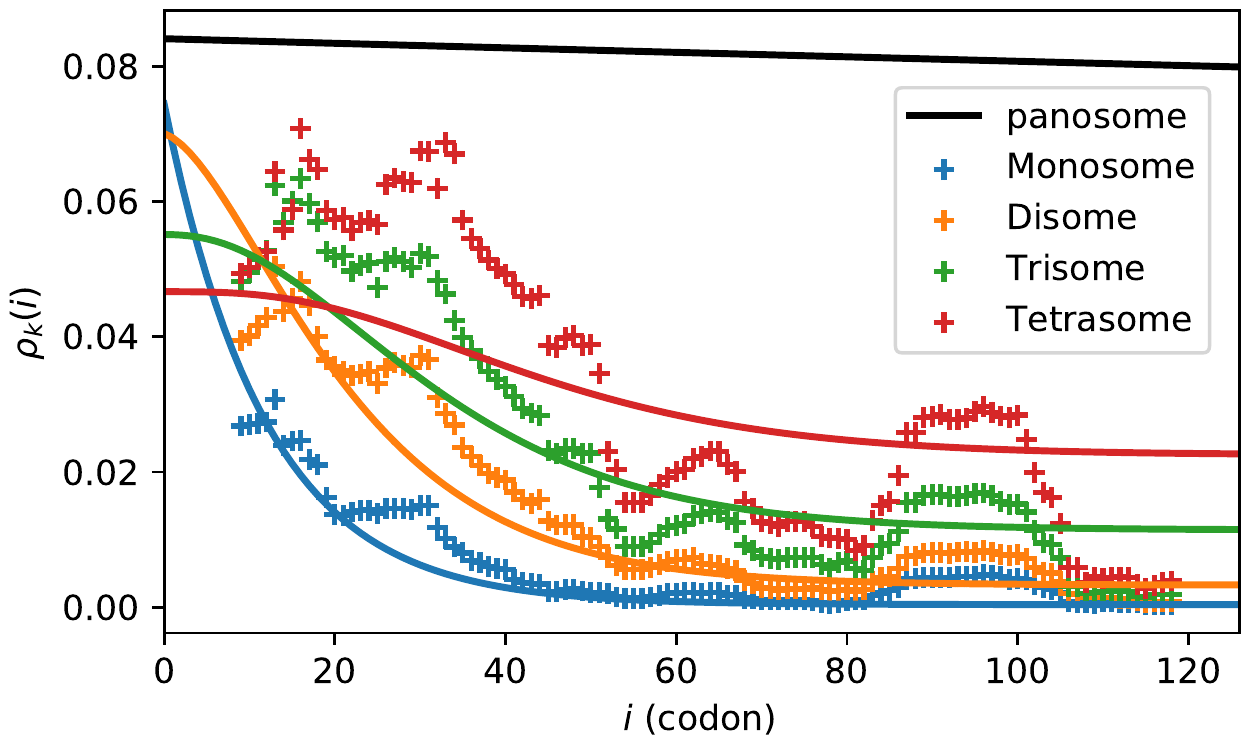} 
    \includegraphics[width=0.9\columnwidth]{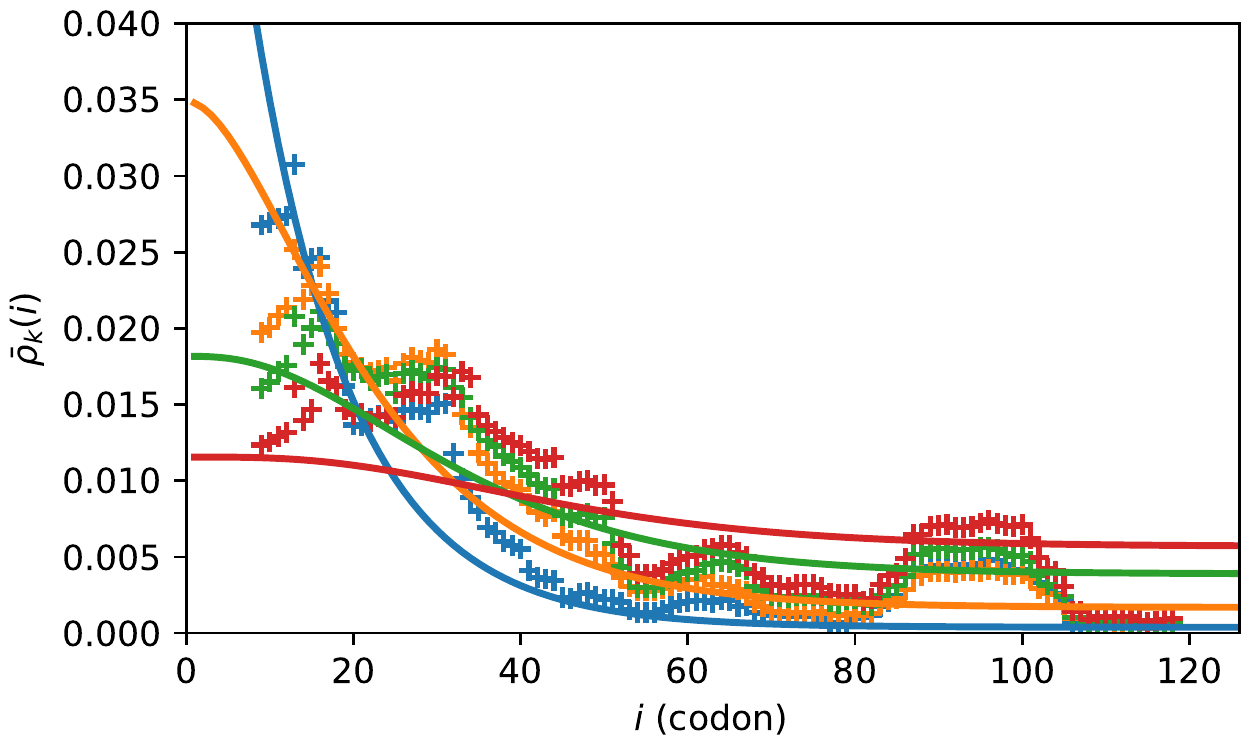}
    \caption{(top) Polysome profiles of a representative experiment. RNA populations were separated on a 15-50\% sucrose gradient under 35,000rpm centrifugation for 3.5h. Free RNA, 40S ribosomal subunit, 60S ribosomal subunit, mRNA charged with 1,2,3\dots ribosomes were separated through a 254 nm UV spectrometer and collected with an ISCO fractionation system. Fractions corresponding to mRNA charged with 1,2,3 or 4 ribosomes were pooled and subjected to RPF extraction and cDNA construction.
    (middle) $k$-some densities $\rho_k$ ($k$=1 to 4) obtained from the homogeneous ballistic model (solid lines) and from  Ribo-seq profiling experiments (symbols) are plotted versus the genomic coordinate (in codon unit). The polysome density is plotted in black for the ballistic model only (lack of experimental data). Experimental data correspond to the histone gene HIST1H2BM and the fit is based on the ballistic model (see text for details) with  fixed parameters, $\alpha=0.06$s$^{-1}$ and $\omega=1/60$ min$^{-1}\approx 3.10^{-4}s^{-1}$. The fit then yields $\mathcal{T}(L)=178$s and hence $p=0.7$ codons/s. 
    (bottom) Same $k$-some densities are normalized to 1, i.e. $\bar{\rho}_k=\rho_k/k$, in order to compare the discrepancies with each-other and show that we retrieve the characteristic crossing between ballistic $k$-somes curves.}
    \label{fig:RiboSeq}
\end{figure}

Figure~\ref{fig:RiboSeq} displays the $k$-some densities, $\rho_k$ (for $k$=1 to 4), versus the genomic coordinate obtained from Ribo-seq profiling experiments for the histone gene HIST1H2BM (symbols) and from a fit of these data using the homogeneous ballistic model (solid lines).

The polysomes were fractionated following the method of~\cite{relier_fto-mediated_2021} in order to separate the mRNAs bound either to one, two, three or four ribosomes. This yields four subsets of mRNA respectively termed mono-, di-, tri-, and tetra-somes (Figure~\ref{fig:RiboSeq}(A)). For each fraction, we extracted all mRNA segments covered by a ribosome (called Ribosome Protected Fragments (RPF)), and prepared a library for sequencing as described in~\cite{gao_genome-wide_2016}. Then short read sequencing was performed as for RNA-seq using an Illumina platform. The sequencing data for each library (mono-, di-, tri-, and tetra-somes), which contains raw sequencing reads, are processed the same way, one independently of the others (using a method adapted from~\cite{paulet_ribo-seq_2017}). For a given mRNA/gene, the goal of bioinformatic processing is to tabulate how many reads cover a given position in the mRNA coding sequence. After cleaning, each read sequence is aligned to the mRNA sequence to determine the position of origin of the RPF – 
a classical task of sequence comparison called mapping – using efficient software~\cite{philippe_crac_2013}. Only reads having a high quality alignment and a unique 
origin location are further processed. One records the matching position of the first base of the read. On the first position of each read, we applied a shift of 12 nucleotides to assign the read to the P-site position of ribosomes~\cite{paulet_ribo-seq_2017}. 
It has been observed from previous analyses that 12 nucleotides is by far the most frequent distance between the read start and the P-site~\cite{stadler_wobble_2011}, which determines the position of the codon being translated. Finally, we record the counts for each P-site covered position of the mRNA in a table.

In the intermediate panel of Fig.~\ref{fig:RiboSeq}, 
Ribo-seq data (cross symbols) have been subsequently smoothed over 19 codons by averaging the signal within a sliding bin and normalized such that the integral of each $k$-some density $\rho_k$ corresponds to $k$. As the genomic coordinate increases, we observe a global exponential-like decay of the densities, a phenomenon predicted by the ballistic model when mRNA degradation is taken into account (see Fig.~\ref{fig:theory1}). Furthermore, another clear signature of mRNA degradation may be observed in the lowest panel of Fig.~\ref{fig:RiboSeq} where we display the Ribo-seq $k$-some densities {\it normalized to unity}. These profiles show clearly that near the entrance of the mRNA (over a genomic distance of 30-40 codons), $k$-some density profiles decrease with increasing $k$ while they increase with $k$ beyond that distance. This crossing of the $k$-some density curves is another prediction of the ballistic model in the presence of mRNA degradation. 

Encouraged by the similarities between $k$-some Ribo-seq data and the corresponding ballistic trends noted above, we shall now try to compare them more quantitatively. We shall do so by restricting this first comparison to the {\it homogeneous} version of the ballistic model, i.e., $p(x)$ constant independent of the codon, and to smoothed experimental data. This choice is partly dictated by its simplicity of implementation, but also and foremost, by the lack of polysome data in the Ribo-seq set available for this histone gene (polysome data prove to be crucial in determining the codon-dependent elongation rates, $p_i$, as we shall explain in the next section). In their absence we choose to fit an average (and thus constant) elongation rate $p$ while fixing the two other parameters, $\alpha$ and $\omega$, to reasonable values found in the literature. To do so, we use the specific half-life of histone mRNA ($\tau_{1/2} =$ 40 min) to obtain (by definition)  $\omega=\ln(2)/\tau_{1/2}
\approx 1/60$ min$^{-1}\approx 3.10^{-4}s^{-1}$ \cite{heintz_regulation_1983}
and use the initiation rate for {\it Homo sapiens} given in \cite{trosemeier_optimizing_2019}:  $\alpha=0.06s^{-1}$. Finally, $p$ is estimated through $\mathcal T(L)=L/p$ in order to optimize the fit for all four studied values of $k$. We find\footnote{This value is the average of ${\cal T}(L)$ values estimated for specific values of $k$: $\mathcal T(L)=147$ s for monosomes, 168 s for disomes, 190 s for trisomes and 206 s for tetrasomes. Given the crudeness of this first approach where the codon-dependency of the elongation rates is not taken into account, the dispersion of these results seems reasonable.} $\mathcal T(L)=178$ s, a value that leads to an average number of ribosomes on the mRNA in the stationary regime $\langle k \rangle =\alpha\mathcal T(L)=0.06 \times 178\approx 10$.  

Overall, we observe that the qualitative trends exhibited by the Ribo-seq data are described by the ballistic model and that the inversion of the $k-$some density profiles near the start of the mRNA gives a clear signature that translation in these experiments is influenced by mRNA degradation.
Finally, in contrast to the $k$-somes, the theoretical density for polysomes (black solid line in the intermediate panel of Fig.~\ref{fig:RiboSeq}),
which has no experimental counterpart here, is nearly flat and barely influenced by degradation:
this is in agreement with the regime of intermediate degradation presented in Fig.~\ref{fig:theory1}. This result provides strong further motivation for the need to split the population of mRNA into $k-$somes in order to increase the sensitivity to mRNA degradation. Note that from the analysis of $f(\tilde\omega)$ done in section \ref{secB}, it was expected that in the present case mRNA degradation would  have little influence on the polysomes density since $\tilde\omega= 3.10^{-4}\times178\approx0.05\ll1$. Finally, we conclude this section by remarking the need to introduce an inhomogeneous $p(x)$ to explain the strong and systematic fluctuations along the profiles that cannot be accounted for by experimental fluctuations. In the absence of polysome data to complete the $k$-some set, we deem it premature to attempt such a fit here (experiments in progress).

\section{Fitting the kinetic rates from ribosome density profiles}

\subsection{Limitations in using solely polysomes to fit  kinetic rates}
Usual Ribo-seq experiments involve polysomes only and their modeling does not account for the mRNA degradation rate, $\omega$.
This is justified inasmuch as, for most living cells,  the polysome density is barely sensitive to degradation and this approximation does not lead to significant errors in the estimation of the elongation rates. Yet, the main problem in estimating the kinetic rates from polysomes alone is that there are more unknowns than equations (or equivalently, more model parameters than experimentally measured quantities). Considering henceforth the inherent discreteness of the mRNA, we shall now use $p_i$ as the elongation rate at site (codon) $i \in \{1,\dots,L\}$ and $\rho_i$ for the polysome density at that site instead of their continuous version. Now, even if we assumed the degradation rate $\omega$ to be known from independent experiments, we would still need to estimate $(L+1)$ parameters: the set of $\{p_i\}$ and the initiation rate $\alpha$. As the polysome density $\rho_i$, on the other hand, only provides $L$ data, we are left with an extra unknown.
    
Another problem in estimating the kinetic parameters is that the number of mRNAs from which the Ribo-seq profile is obtained is not known, which prevents a proper normalization of the Ribo-seq density. A parametric estimation must therefore rely on a density $\bar \rho(x)$ normalized to unity, $\int_0^L\bar\rho(x)dx=1$, instead of $\langle k \rangle$. This density is given by 
    \begin{equation} \label{barrhox}
        \bar{\rho}(x)=\frac{\omega}{p(x)}\frac{e^{-\omega \mathcal{T}(x)}}{1-e^{-\omega \mathcal{T}(L)}}.
    \end{equation}
Although this  density no longer depends on $\alpha$,  this does not solve the problem,  since the new normalisation constraint, whose discrete version is $\sum_{i=1}^L \bar{\rho}(i)=1$, also removes one equation for the input parameters. A strategy found in the literature to circumvent this problem is to normalize Ribo-seq profiles by a mean ribosome velocity measured from independent experiments \cite{duc_impact_2018}. This method suffers from several drawbacks, however: (i) the ribosome velocity that is used is averaged over an entire genome instead of being gene specific, (ii) results are typically obtained for species different from the one involved in the Ribo-seq experiment and (iii) experiments are done independently in different conditions. Altogether, the normalization obtained from this procedure is expected to be subject to large uncertainties.

\subsection{$k$-some improved fit of kinetic rates}
Half-lives ($\tau_{1/2}$) of individual mRNAs have been successfully measured in mouse embryonic stem cells \cite{sharova_database_2009}. By applying the same method, any mRNA half-life can be measured as well. We shall 
henceforth consider $\omega = \ln(2)/\tau_{1/2}$ as a known kinetic time scale. The normalised polysome density does not depend on $\alpha$ (Eq. \ref{barrhox}), but on the  $p_i$'s only and,
as we have shown, it is not strongly affected by mRNA degradation. It therefore constitutes a suitable quantity to determine the elongation rates, $p_i$, independently of the other parameters. Furthermore, in section \ref{sec:ksome_density_analysis}, 
we have shown that a $k$-some Ribo-seq data analysis provides additional information on the kinetic parameters, especially when the mRNA lifetime decreases for, in that case, $k$-some densities become less and less linearly related.
Yet, when the mRNA degradation becomes too important (large $\omega$), $k$-some density profiles vanish near the end of the mRNA rendering their analysis quite difficult. The same is true of polysome densities. This, however, happens for parameters far outside the biological range.

In a first approach, we therefore decide to combine the information provided by both the polysome normalized density and the monosome density since among $k$-some profiles, it is the most sensitive to the mRNA lifetime (see section \ref{sec:ksome_density_analysis} and SI, section 5). 

\subsection{Optimizing the fit with respect to the degradation rate}

\begin{figure}[h!]
    \centering
    \includegraphics[width=1\columnwidth]{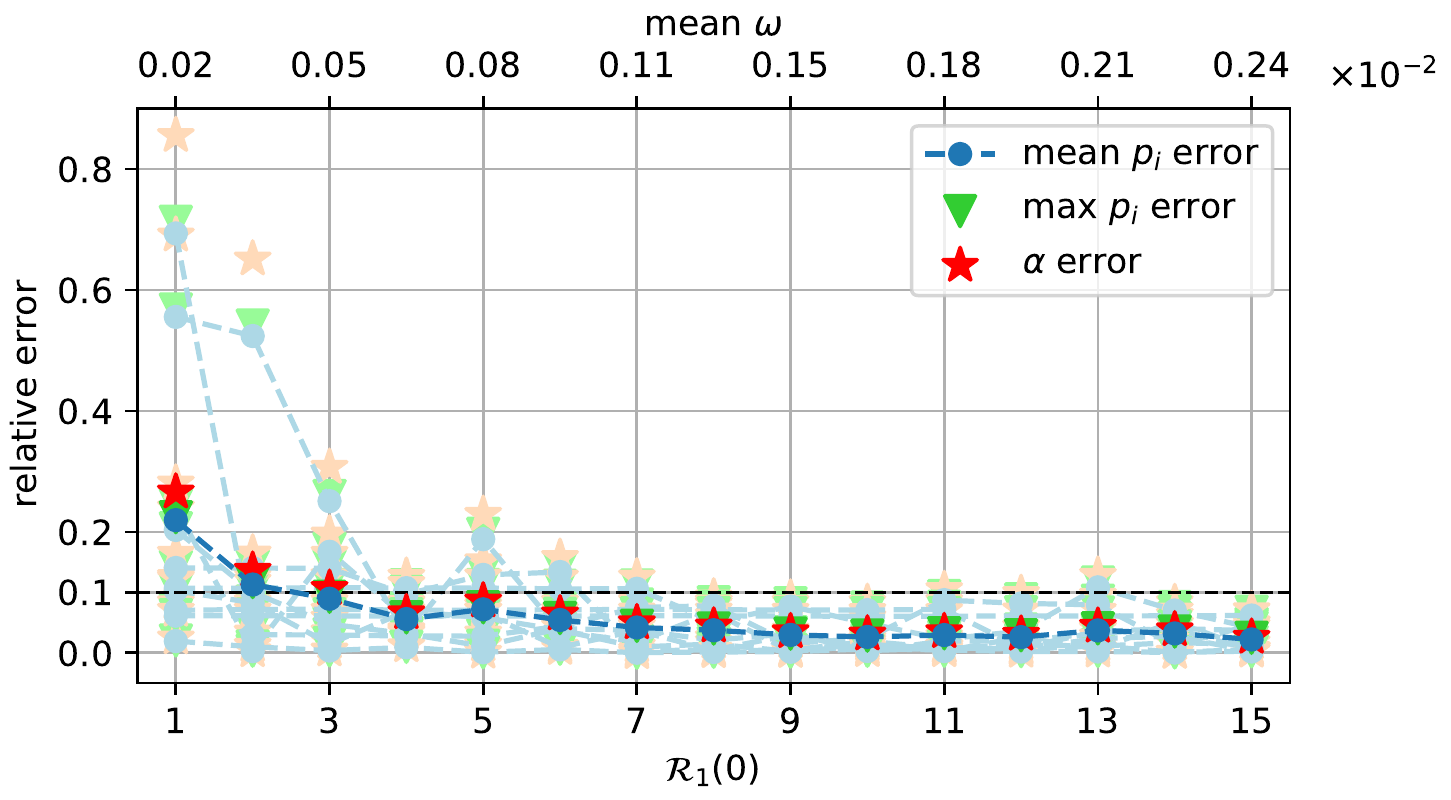}
    \caption{Score functions \eqref{eq:Err} of the fit with the ballistic model when $\alpha=0.08$s$^{-1}$ and for 10 ballistic density profiles obtained by a random drawing from a uniform distribution of the elongation rates $p_i\in[0.2,4[$ s$^{-1}$ (pale colors). Results averaged over the 10 profiles are displayed in vivid colors. $\mathcal{R}_1(0)$, \eqref{ratio1}, is increased by increasing $\omega$ and $\mathcal{T}(L)\simeq 81.3$ s on average.}
    \label{score_fct}
\end{figure}

We now implement the strategy outlined above by detailing how to fit  the initiation rate $\alpha$ and the elongation rates $p_i$ of all the codons constituting the mRNA. To find these kinetic parameters, we minimize the quadratic difference between the ballistic model predictions and, simultaneously, the Ribo-seq data for the normalized (to one) polysome density and the monosome density. To be more specific, we minimise the quantity
\begin{equation}
\mathbb{X}^2=\chi_{polysome}^2+\chi_{1-some}^2
\label{eq:chi}
\end{equation}
where $\chi_{polysome}^2$ is the sum of the relative quadratic errors of the normalized polysome density on each codon :
\begin{equation}
\chi_{polysome}^2=\sum_{i=1}^L \left|\frac{\rho_i^{data}-\rho_i^{ballistic}}{\rho_i^{data}}\right|^2,
\end{equation}
and where $\chi_{1-some}^2$ has a similar form for monosomes. The minimisation, performed by the L-BFGS algorithm  (an adaptation of the  Broyden-Fletcher-Goldfarb-Shanno quasi-Newton method for using a limited amount of computer memory) encoded in the Python 3 library Scipy  \cite{noauthor_scipyoptimizeminimize_nodate}, yields at once $\alpha$ and the $L$ parameters $p_i$.

As we shall see, the accuracy of our fitting method improves when the values of the ratio 
\begin{equation} \label{ratio1}
\mathcal R_1(0)=\Tilde{\omega}\left(e^{\Tilde{\omega}+\Tilde{\alpha}}-1\right)/(\Tilde{\omega}+\Tilde{\alpha})    
\end{equation}
(see \eqref{ratiok} and SI section 5) increase. As a first feasibility test, the minimisation procedure is performed on artificial data created by the ballistic model itself. The goal is to measure its accuracy in recovering the known input parameters as a function of $\mathcal R_1(0)$.
The relative error between a parameter value found by minimisation, $Y^*$, and its exact value, $Y$, is defined by
\begin{equation}
    {\rm Err}_Y=\frac{|Y-Y^*|}{Y}\,.
    \label{eq:Err}
\end{equation} 
We call it a "score function". Beside ${\rm Err}_{\alpha}$, and to avoid the wealth of score functions related to every single elongation rate $p_i$, we shall estimate two quantities: an average score function ${\rm Err}_{\rm mean} = \sum_i {\rm Err}_{p_i}/L$, and the maximum of them all, ${\rm Err}_{\rm max} = \max_i {\rm Err}_{p_i}$. 

To test our minimisation procedure, we fix $\alpha=0.08$ s$^{-1}$. We then generate ten sets of $L$ random $p_i$ values drawn from a uniform distribution in the range $[0.2,4]$ s$^{-1}$. Each set has its own $\mathcal T(L)=\sum_i p_i^{-1}$ value and $\omega$ is adjusted in such a way that the ratio $\mathcal R_1(0)$ has a desired value (here, an integer from 1 to 15). Now, to each set of $\alpha$, $p_i$ and $\omega$ values correspond ballistic polysome and monosome densities. These densities are the artificial data processed by minimisation of the $\mathbb{X}^2$ function \eqref{eq:chi} as explained above. This procedure yields the parameter values $\alpha^*$ and $p_i^*$ from which the score functions ${\rm Err}_{\alpha}$, ${\rm Err}_{\rm mean}$ and ${\rm Err}_{\rm max}$ are then evaluated to check the accuracy of the method.

Figure \ref{score_fct} displays the 3 score functions for the ten density profiles sharing the same $\alpha=0.08$ s$^{-1}$ and the same $\mathcal R_1(0)$ (pale colors). The observed variability stems from the randomness of the $p_i$ sets. To smooth out this variability, score functions have subsequently been averaged over the ten random profiles (vivid colors). We can see that relative errors decrease when $\mathcal R_1(0)$ increases: they are less than $10\%$ for $\mathcal R_1(0)\geq 3$ and remain nearly constant (around 1\%) for $\mathcal R_1(0)\geq 10$. This confirms the relevance of $\mathcal R_1(0)$ as a good criterion for the fit quality, albeit in a suitable parameter range. In the presented fits, $\mathcal R_1(0)=3$ corresponds to the parameters $\alpha=0.08$ s$^{-1}$, $\mathcal{T}(L)\approx 81$ s and $\omega\approx 0.0005$ s$^{-1}$ $\approx \frac{1}{33}$ min$^{-1}$, which are all biologically reasonable rates for mRNA translation. This analysis therefore allows us to assess the feasibility of our proposed fitting method to determine the kinetic parameters of mRNA translation for a particular gene.

\section{Translation \textit{phase diagram}: the  effect of finite mRNA lifetime}

\begin{figure}[h!]
    \centering
    \includegraphics[width=1.0\columnwidth]{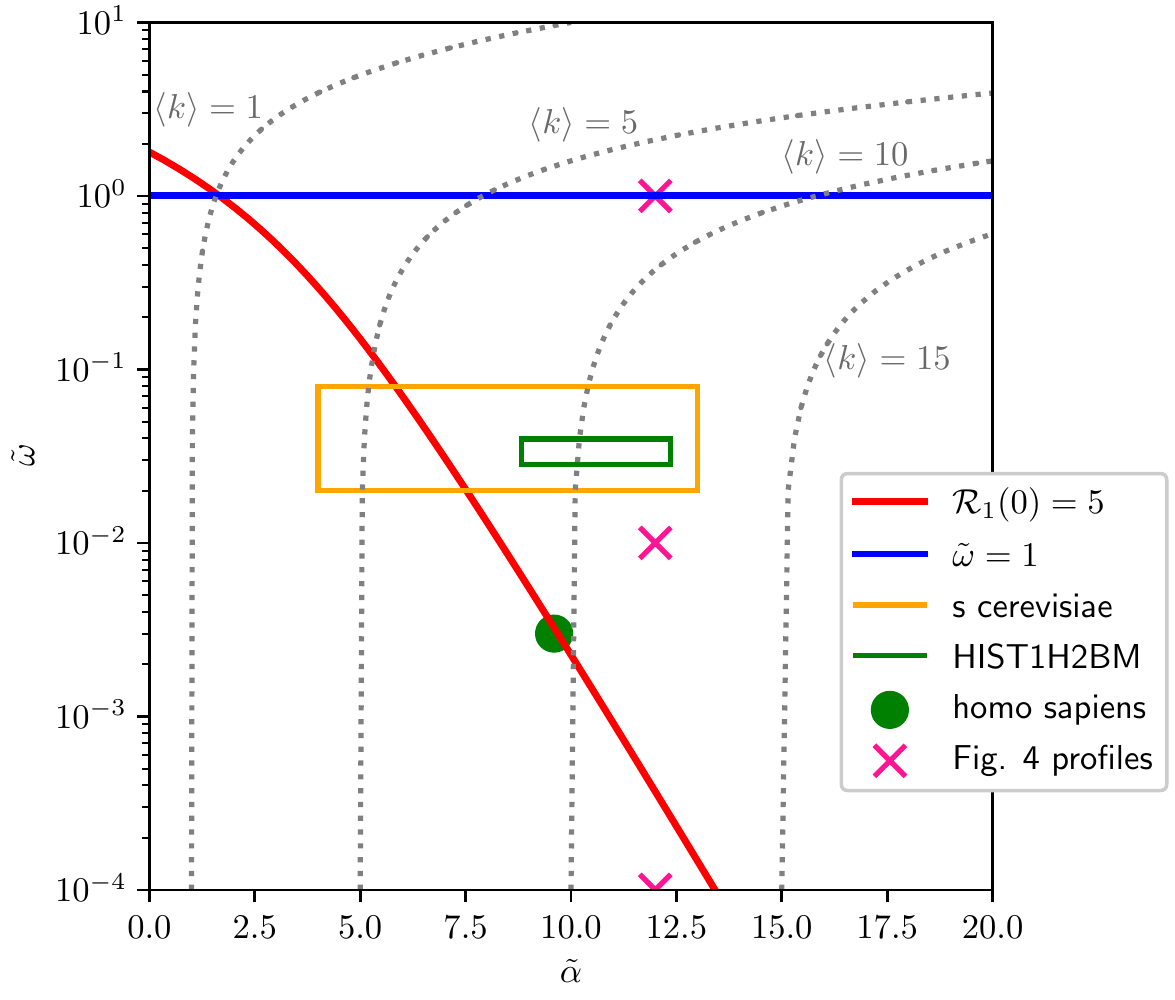}
    \put(-150,195){\large High Degradation (HD)}
    \put(-175,150){\large Intermediate Degradation (ID)}
    \put(-210,30){\large Low Degradation (LD)}
    \caption{Phase diagram of the ballistic model in the plane $(\tilde\alpha,\tilde\omega)$. The phase diagram is split in three degradation regimes with the lines of equation $\tilde\omega=1\sim\mathcal O(1)$ and $\mathcal R_1(0)=5\sim\mathcal O(1)$. 
    Biological values obtained for the histone HIST1H2BM (human) are given by the homogeneous fitting from monosome to tetrasome (Fig. \ref{fig:RiboSeq}). 
    }
    \label{fig:fit_diag}
\end{figure}

We summarize our main results in Fig.~\ref{fig:fit_diag}. This figure can be considered as a {\it phase diagram} of the ballistic model in the  $(\tilde\alpha,\tilde\omega)$, or (initiation, degradation),  plane. It indicates the three phases already defined in section \ref{sec:ksome_density_analysis}: low degradation (LD), where neither polysome nor $k$-some ribosome density profiles are affected by the finite mRNA lifetime, intermediate degradation (ID), where only $k$-some densities (with small $k$) are affected, and high degradation (HD), where both polysome and $k$-some densities are impacted. 
The LD and ID regimes are separated by the curve $\mathcal R_1(0) =5$, above which $k$-somes feel a pronounced effect of degradation. The ID and HD regimes are separated by the line $\tilde\omega=1$, where polysomes are strongly affected by degradation, which becomes visible on the constant $\langle k \rangle$ curves in the ID phase and therefore on  polysomes when $\tilde\omega > 0.1$, see also Fig.~\ref{fig:k_av}. 

The ID phase also turns out to be the most efficient regime for our fitting method.
In this regime, the equations for the $k$-some densities are decoupled and the impact of degradation is not yet strong enough for $k$-some densities to be too low at the end of the mRNA transcript to obtain information about $p_i$ variations there.
We have noted however that the quality of the fits doesn't decrease until  $\tilde\omega = 5$ is reached, where the polysome density become of the order of $10^{-5}$ near the end of the mRNA and information on the corresponding elongation rates become undetectable.

The dashed lines represent curves of constant $\langle k\rangle$ 
(Eq.(\ref{k_av}))  
in the  $(\tilde\alpha,\tilde\omega)$  plane. As degradation increases, we observe a departure from the value $\langle k \rangle_\infty =\tilde\alpha$ that is a good approximation in the intermediate regime. We also see that this value does not evolve substantially until the boundary between the ID and HD phases is approached. We observe that the line $\langle k \rangle=1$ intersects the curves $\mathcal R_1(0) = 5$ and $\tilde\omega=1$ at the same point. An average value $\langle k \rangle=1$ is then at the limit of feasibility for our proposed fitting method.

We have evidence from the results shown in Fig.~\ref{fig:fit_diag} that at least some histonic genes (green rectangle), as well as the average yeast genome value ({\it S. cerevis\ae}, yellow rectangle), fall in the ID phase. The average for the {\it H. sapiens} genome, however, appears to fall close to the boundary between the LD and ID phase (red line) and may be at the limit of feasibility of our fitting method. However, our analysis do suggest other routes to  determining the model parameters, e.g. by fitting the k-some repartition $P_k$ (\eqref{Pkn}), an experimentally accessible quantity.

It is interesting to note that in the ID regime, the method we propose to extract the kinetic parameters of translation exploits a monosome density profile that is experimentally accessible despite the possibly much lower probability of finding a monosome than a $k$-some with $k \approx \langle k\rangle_\infty \approx \tilde\alpha$ ($P_1 \ll P_{\tilde\alpha}^\infty$). In that sense, a "deep" sequencing (billions of mRNAs corresponding to millions of cells) is necessary to study monosomes as thousands of them are needed in order to ensure a statistically relevant analysis (see discussion in section \ref{sec:ksome_density_analysis}).

\section{Discussion and Conclusions} 

In this paper, we present an experimental 
setup where mRNAs are separated, before being sequenced, according to the number of ribosomes attached to them. In this context, and for the first time to our knowledge, we develop a ballistic model of ribosome transport tailored to the analysis of such Ribo-seq experiments in the presence of mRNA degradation.

By investigating the  mean ribosome number $\langle k\rangle$, as well as polysome and $k-$some densities, we characterize the system by identifying three regimes that emerge from a  parametric analysis of the ballistic model. Provided that $\langle k\rangle_\infty = \tilde\alpha > 1$,  we arrive at the following global picture: in the low degradation (LD) regime the effect of the finite mRNA lifetime on both $\langle k\rangle$ and the densities is negligible; in the  intermediate degradation (ID) regime, the finite mRNA lifetime has a strong impact on $k-$somes densities, but not on polysome densities (and therefore little impact on  $\langle k\rangle$); in the   high degradation (HD) regime, all densities, as well as  $\langle k\rangle$, are affected by the finite mRNA lifetime. The HD regime is delimited by $\tilde{\omega}=1$ for which the time for a ribosome to cross the mRNA, $\mathcal{T}(L)$, is equal to the inverse of the degradation rate $\omega$. This limit probably does not correspond to any usual biological regime,  which leads us to the conclusion that polysome densities should probably never be affected by finite mRNA lifetime (except perhaps in rare extreme conditions).

Given that the ID regime is the biologically relevant one, we can base a viable fitting strategy on the monosome and normalized polysome densities alone.  Because the finite lifetime strongly affects the monosome density, and only slightly the polysome one, the two densities give access to complementary information. Whereas the monosome density exhibits the typical shape expected from finite mRNA lifetime (exponentially decreasing profiles) depending on $\alpha,~\omega$ and $\mathcal{T}(L)$, the normalized  polysome density is mainly sensitive to $p_i$ variations. Precisely, the polysome density no more depend on $\alpha$ when being normalized, and as $\omega$ is measured by another experiment, unknowns are only $p_i$.
Furthermore, we  have identified a quantity directly related to the effectiveness of the method, namely the ratio of transient to stationary monosome densities, $\mathcal R_1(0)$,  \eqref{ratio1}. 
Since the experiment presented in this article involve measuring the density profiles for both the polysomes and the $k$-somes with $k = 1,\ldots,4$, the model predictions for the disomes, trisomes, and tetrasomes (not used in the fitting procedure) provide a test of the consistency of the modeling approach we propose in ID regime.

Using our ballistic model, we have analyzed preliminary Ribo-seq data for a human histonic gene. It is clear from this analysis that the Ribo-seq data display the same phenomenology predicted by the ballistic model in the ID regime: a decrease of the $k-$some Ribo-seq profiles at the end of the mRNA due to the degradation of mRNA in the late stages of translation. Thus we show for the first time the effect of degradation on Ribo-seq data. 

In Fig.~\ref{fig:fit_diag}, we present a phase diagram of the ballistic model in the $(\tilde\alpha,\tilde\omega)$ plane as a summary of the main results of the finite lifetime effect study and the fitting method. This phase diagram displays the three regimes: LD, ID and HD, the ID regime considered as the biological regime as well as the regime where our fitting method is effective. 

In the modeling of real experimental data it will also be important to evaluate the influence of experimental noise on the reliability of our fitting method. It would be interesting to look at the effect of degradation with finite resources of ribosomes~\cite{ciandrini2014motor} and finite diffusion of  ribosomes~\cite{dauloudet2021modelling,dauloudet2021erratum}.

\bibliography{mainTMP}
\end{document}